# Vector magneto-optical magnetometer based on the resonant all-dielectric gratings with highly anisotropic iron-garnet films

D.O. Ignatyeva[1,2,3], G.A. Knyazev[1,3], A.N. Kalish[1,3], A.I. Chernov[1,2,4],
V.I. Belotelov[1,2,3]

[1] Russian Quantum Center, Skolkovo Innovation City, 30 Bolshoy Bulvar, Moscow 121353, Russia

[2] Vernadsky Crimean Federal University, 4 Vernadskogo Prospekt, Simferopol 295007, Russia

[3] Faculty of Physics, Lomonosov Moscow State University, Leninskie Gory, Moscow 119991, Russia

[4] Center for Photonics and 2D Materials, Moscow Institute of Physics and Technology, National Research University, 9 Institutskiy per., Dolgoprudny 141700, Russia

A sensitive vector magnetometry with high spatial resolution is important for various practical applications, such as magnetocardiography, magnetoencephalography, explosive materials detection and many others. We propose a magnetometer based on the magnetic iron-garnet film possessing a very high magnetic anisotropy, placed in the rotating external magnetic field. Each of the measured magnetic field spatial components produces different temporal harmonics in the out-of-plane magnetization dependence. The dielectric resonant grating placed on the top of an ultrathin film enhanced the magneto-optical response 10 times which makes it possible to achieve 10 times higher spatial resolution in the perpendicular to the film direction. The reported magneto-optical magnetometer allows one to measure simultaneously all three spatial components of the magnetic field with high spatial resolution and sensitivity up to 100 pT/Hz$^{1/2}$.

## 1. Introduction

Registration of weak magnetic fields is an important tool for study of different animate and inanimate objects. Magnetocardiography based on the measurement of the small magnetic signals associated with the blood currents provides novel information essential for diagnostics [1-2]. Detection of weak magnetic signals is important, for example, for magnetoencephalography [3], sensing of explosive materials [4], particle physics [5].

For many practical applications, it is essential to provide not only high sensitivity of the magnetometer, but also micrometer spatial resolution and an ability to distinguish orientation of the external magnetic field. There are many types of the magnetometers developed nowadays, among which one of the most perspective approaches is the flux-gate magnetometry [6−9] reaching a 100 fT/Hz$^{1/2}$ sensitivity [9] in a wide frequency range up to GHz frequencies. The flux-gate magnetometry is based on the measurement of the medium magnetization under the impact of the external magnetic field. It was recently shown that contactless magneto-optical reading of the magnetization state allows one to measure magnetic fields in vector format with micrometer spatial resolution. Therefore, the enhancement of the magneto-optical effects is a vital problem of the flux-gate magnetometry.

Recently various nanostructures were demonstrated to enhance the magneto-optical effects, among which are photonic crystal-based [10-13], nanoplasmonic [14-16] and all-dielectric ones [17-20]. Due to the low optical losses and simplicity of fabrication and measurements, all-dielectric nanostructured films seem to be one of the most promising.

We consider a 1D $TiO_2$ grating deposited on the top of the smooth ultrathin iron-garnet film grown on a gadolinium-gallium garnet (GGG) substrate and having a very high magnetic anisotropy. Performed analysis and numerical simulations show that one can perform vector magnetometry in the proposed structure with a high sensitivity up to 100 pT for Hz bandwidth. The subwavelength size of grating perforation and film width allows for the significant device miniaturization. Actually, the magnetic film thickness determines the spatial resolution of the magnetometer which is 30 nm in the reported device.

## 2. Dynamics of magnetization in magnetic film with strong anisotropy

Let us consider a magnetic dielectric film of cubic crystal lattice and crystallographic axis orientation (111) placed into an external saturating magnetic field **H** (see Fig. 1(a,b)) that uniformly rotates in the film plane with a frequency $\omega$. The problem also contains a weak measurable field $\mathbf{h} \ll \mathbf{H}$ with an arbitrary direction and thus having arbitrary $\mathbf{h} = \{h_x, h_y, h_z\}$ spatial components. [21]. We consider a film with high cubic anisotropy, for example $Tm_3Fe_{4.3}Sc_{0.7}O_{12}$ [22] or $(Bi_{1.1}Lu_{1.45}Pr_{0.2}Tm_{0.2}Gd_{0.05})(Fe_{3.5}Al_{0.8}Ga_{0.7})O_{12}$ [23]. The dynamics of the magnetization **M** vector under the **H** impact is rather complicated in such films, moreover, **M** direction does not coincide with **H** even in the quasistatic case ($\omega \ll \gamma H$). Let us now discuss how this complicated dynamics gives rise to the 3D magnetometry and allows the BIG magnetization to be sensitive to all of the **h** spatial components.

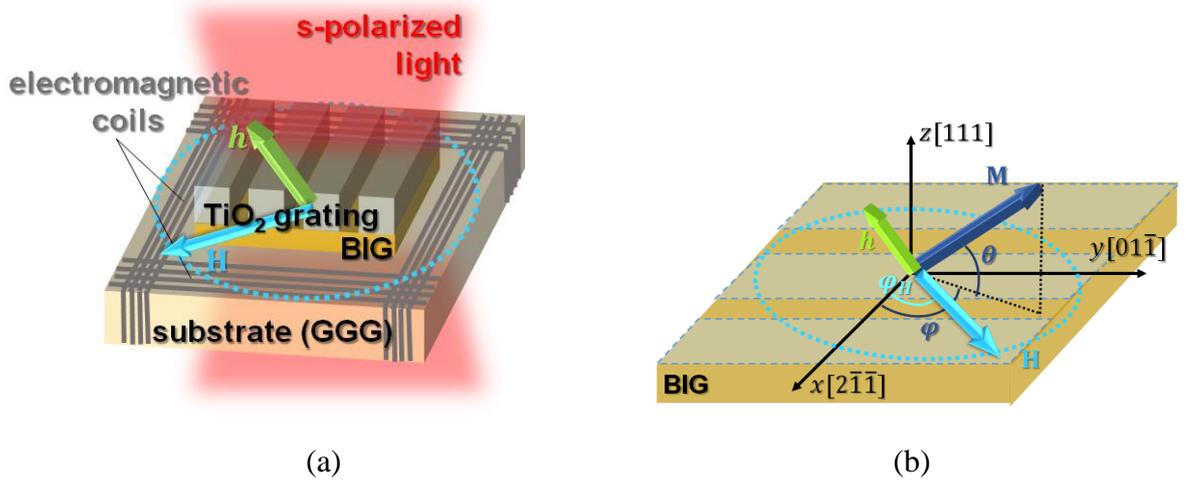

(a) (b)

Figure 1. 1D grating-based magneto-optical vector magnetometer. (a) Device principal scheme. (b) Magnetization **M** motion in the BIG film with high magnetic anisotropy under the action of the rotating magnetic field **H**.

The condition for the minimum free energy of a magnetic film gives the following expressions for the out-of-plane angle $\theta$ and azimuthal angle $\varphi$ (see Fig. 1(b)) that determine the direction of the magnetization in the quasi-static case [23]:

$$\theta = \frac{\frac{\sqrt{2}}{3}K_1 \sin(3(\varphi-\varphi_K))+h_z M_s}{4\pi M_s^2 - 2K_U - K_1 + HM_s \cos(\varphi-\varphi_H) + M_s h_x \cos\varphi + M_s h_y \sin\varphi}, \tag{1}$$

$$M_s h_x \sin\varphi - M_s h_y \cos\varphi + HM_s \sin(\varphi - \varphi_H) - \sqrt{2}\theta K_1 \cos(3(\varphi - \varphi_K)) = 0, \tag{2}$$

where $\varphi_K$ is the angle between the anisotropy axis $[2\bar{1}\bar{1}]$ and the axis $X$, $\varphi_H$ is the angle between **H** and $X$, $K_1$ is the cubic anisotropy constant, $K_U$ is the uniaxial anisotropy constant (see Fig. 1(b) scheme). The analytical description is performed for arbitrary values of $\varphi_K$, while in numerical simulations $\varphi_K = 0$ for the definiteness.

Since we consider a rotating magnetic field, the following condition describes magnetic field direction:

$$\varphi_H = \omega t + \varphi_K, \tag{3}$$

$$|H| = const. \tag{4}$$

Analysis shows that under the assumption that the measured field is small ($H \gg h$) we can obtain the following equations for $\theta$ and $\varphi$:

$$\theta = \frac{K}{A}\left(\frac{1}{3}\sin(3\omega t) + \frac{h_y}{2H}(\cos(4\omega t + \varphi_K) + \cos(2\omega t - \varphi_K)) + \frac{h_x}{2H}(\sin(2\omega t - \varphi_K) - \sin(4\omega t + \varphi_K)) + \frac{h_z K}{2HA}(1 + \cos(6\omega t))\right), \tag{5}$$

$$\varphi = \frac{h_y \cos(\omega t + \varphi_K) - h_x \sin(\omega t + \varphi_K) + \frac{K^2 \sin(6\omega t)}{6A} + \frac{h_z}{A}K\cos(3(\omega t))}{H} + \omega t + \varphi_K. \tag{6}$$

In Eqs. (5) and (6), we use the following designations:

$$A = 4\pi M_s - 2K_U/M_s - K_1/M_s + H, \tag{7}$$

$$K = \frac{\sqrt{2}K_1}{M_s}.$$

Fourier spectra of the magneto-optical signal associated with $\theta(t)$ determined by Eq. (5) provides different temporal harmonics corresponding to the external and measured magnetic field components:

$$\theta_{3\omega} = \frac{K}{3A}\sin(3\omega t)$$

$$\theta_{4\omega} = \frac{K}{2AH}h_\tau \cos(4\omega t + \varphi_K + \varphi_h)$$

$$\theta_{2\omega} = \frac{K}{2AH}h_\tau \cos(2\omega t - \varphi_K - \varphi_h) \tag{8}$$

$$\theta_{6\omega} = \frac{K}{2AH}h_z \cos(6\omega t),$$

where $h_\tau = \sqrt{h_x^2 + h_y^2}$, and $\varphi_h = \tan^{-1}\frac{h_x}{h_y}$. Therefore, according to Eq. (8), measurement of only $\theta(t)$ dependence is enough to determine both in-plane and out-of-plane components of the weak magnetic field **h** via the magnitudes and the phases of the corresponding harmonics. Typical values of $K$=6.7 Oe and $A$=1.6 $10^3$ Oe [22] in the iron-garnets produce rather small values of $\theta_{3\omega} \approx$ 0.01. Therefore, amplification of the observed $\theta$-associated magneto-optical signal is important for the increase of the magnetometer sensitivity. Taking thicker iron-garnet films [23] allows one to increase the Faraday rotation angle, but simultaneously lowers the spatial resolution in z-direction. We propose an alternative approach based on the fabrication of the non-magnetic nanostructure on the top of an ultrathin iron-garnet film that amplifies the magneto-optical response due to the excitation of the optical resonances.

## 3. Magneto-optical magnetic field sensing in 1D gratings

Magneto-optics is a very convenient tool for measurement of the out-of-plane magnetization component of the iron-garnet film, which contains the information on both in-plane and out-of-plane components of the measured weak magnetic field. There are two magneto-optical effects, the polarization Faraday rotation and the even magneto-optical intensity modulation, namely, which arise from the $M_z$ component and should be taken into account. Both of these effects can be significantly enhanced in nanostructures contacting iron-garnet films compared to the smooth materials.

We consider the s-polarized light incident normally on the nanostructure designed in the following way. 1D TiO$_2$ grating having period $P = 370$ nm, TiO$_2$ stripe width $W = 335$ nm, and height $d_{gr} = 240$ nm is fabricated on a top of an ultrathin iron-garnet film with thickness $d_{BIG} = 30$ nm grown on a gadolinium gallium garnet (GGG) substrate. Such 1D grating is resonant, which means that no propagating diffraction orders in the reflectance or the transmittance are generated. On the other hand, it has a narrow resonance at $\lambda = 783$ nm corresponding to a TE0-guided mode excited in the hybrid iron-garnet+TiO$_2$ grating layer. We have tuned the parameters of the grating to make $\lambda$ correspond to the typical wavelength of laser diode, however it can be easily tuned to any desired wavelength. The exact resonance position is determined by the period of the grating and the guided mode effective refractive index $n_\beta$ as $\lambda_{res} = n_\beta P$ [19].

The optical transmittance spectra and the spectra of the Faraday rotation angle for a smooth iron-garnet film and a heterostructure with resonant grating are shown in Fig. 2(a). Notice a two orders enhancement of the Faraday rotation angle compared to the smooth film exhibiting Faraday rotation equal to $\Phi = 0.04°$. This enhancement is spectrally close to the transmittance gap corresponding to the guided mode excitation. Therefore, for possible practical applications, one should bear in mind also the magneto-optical figure of merit MO FOM determined as $MO\ FOM = 0.1 \cdot \log 10 \cdot (\Phi/\log T^{-1})$ [deg/dB], where $T$ is transmittance. From the point of view of the increase of the sensitivity of the magnetometer, it is important to increase signal-to-noise ratio. The detected signal is determined by the Faraday rotation $\Phi$ and the noise is determined by the fluctuations of the laser intensity transmitted through the structure and characterized by the $\sqrt{\Delta T/T}$ magnitude [23]. Therefore, a value of the magneto-optical signal determined as $MO\ signal = \Phi\sqrt{T}$ will also be considered. Actually, the sensitivity of the sensor defined as the minimal registered value of the magnetic field $h_{min}$ is defined by the following equation:

$$\frac{h_{min}}{\sqrt{\Delta f}} = \frac{2AH}{K\Phi_0\sqrt{T}}\sqrt{2\frac{\hbar\omega_\lambda}{\mu P}}, \tag{9}$$

where $\Delta f$ is the spectral bandwidth of the sensor, $\mu$ is the quantum yield of the photodetector, $P$ is the power of the incident light, $\hbar\omega_\lambda$ is the energy of photons of incident radiation.

Let us also notice that the optical resonance is very narrow: its spectral width is $\Delta\lambda = 6.5$ nm. The resonance of the magneto-optical Faraday rotation is about 2.5 times narrower. In order to take into account, on the one hand, very sharp resonances, and on the other hand, the finite spectral width of the light source, we performed an averaging of the values *T*, Φ, *MO FOM*, *MO signal* over the finite bandwidth ($\Delta\lambda_{av}$ =0.2 nm was taken as the typical spectral width of the laser diode radiation). Such consideration did not significantly affect the characteristics of the structure.

MO FOM and MO signal were calculated for the considered structure. Fig. 2© shows that in spite of the transmittance decrease, both quantities in the considered nanostructure are more than one order higher than in a smooth uncoated film. Indeed, for a wavelength $\lambda = 782.5$ nm the transmittance is $T = 5\%$, the Faraday rotation is $\Phi = 1.6$ deg so the *MO FOM*=0.1 deg/dB and *MO signal*=0.16 deg are more than 10 times higher than *MO FOM*=0.016 deg/dB and *MO signal*=0.009 deg values for the smooth uncoated film.

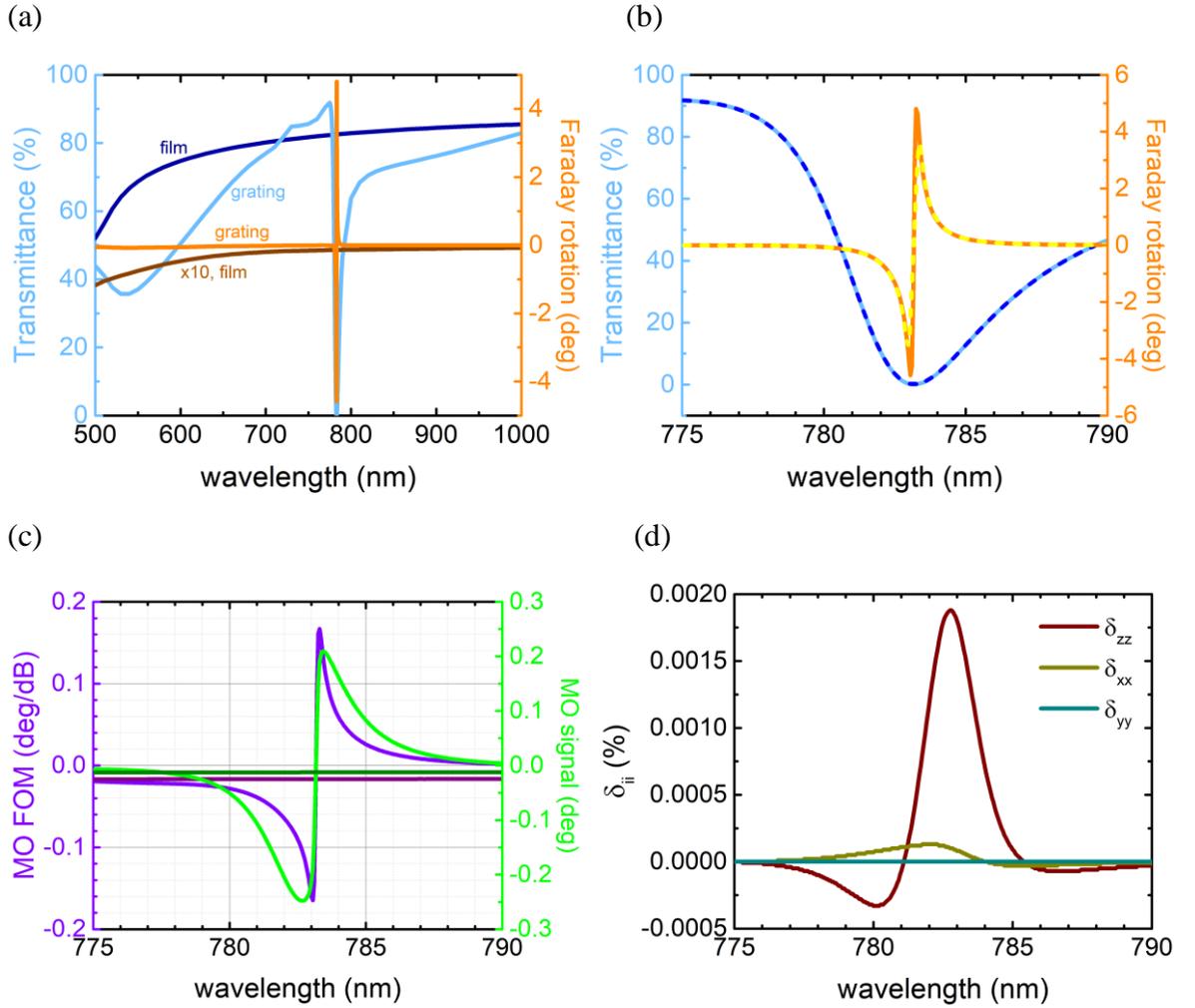

Figure 2. Optical and magneto-optical spectra of the 1D resonant iron-garnet grating. (a) Spectra of the transmittance (blue) and the Faraday rotation (orange) of the uncoated iron-garnet film (dark lines) vs. spectra of the iron-garnet film with a dielectric grating (light lines), see parameters in the text. (b) Spectra of the transmittance (blue) and the Faraday rotation (orange) in the magnified wavelength scale: (solid lines) calculated for a certain $\lambda$ (dashed lines) averaged over $\Delta\lambda_{avg} = 0.2$ nm. (c) MO FOM (violet) and MO signal (green) of the uncoated iron-garnet film (dark lines) vs. spectra of the iron-garnet film with a dielectric grating (light lines), averaged over $\Delta\lambda_{avg} = 0.2$ nm bandwidth. (d) Transmittance modulation coefficients $\delta_{ii}$ calculated for the considered structure.

The Faraday polarization rotation $\Phi$ of the transmitted light is directly proportional to the out-of-plane magnetization component determined by the value $M_z \propto \sin\theta \approx \theta$. One may assume $\Phi = \Phi_0 \theta$ where $\Phi_0$ is the Faraday rotation angle corresponding to the sample magnetized up to saturation $M_z = M_s$ and $\theta(t)$ is determined by the Eq. (5). The weak magnetic field **h** and external field **H** components could be measured from the following harmonics of $\Phi(t)$:

$$\Phi_H = \Phi_0 \frac{K}{3A} \sin(3\omega t),$$

$$\Phi_\tau = \Phi_0 h_\tau \frac{K}{2AH} (\cos(4\omega t + \varphi_K + \varphi_h) + \cos(2\omega t - \varphi_K - \varphi_h)),$$

$$\Phi_z = \Phi_0 \frac{K}{2AH} h_z \cos(6\omega t), \tag{10}$$

so that the signal at the frequency $3\omega$ corresponds to the applied magnetic field, the signal at $2\omega$ and $4\omega$ correspond to the in-plane component of the measured magnetic field, and the signal at $6\omega$ is responsible for the out-of-plane component of the measured magnetic field. The phase of the signal at $2\omega$ and $4\omega$ frequencies carries the information on the in-plane direction of the measured magnetic field $\varphi_h$.

Fig. 3 illustrates the different contributions of the in-plane and out-of-plane components in the $\theta(t)$ temporal dependence and the numerically simulated optical Faraday rotation in 1D grating.

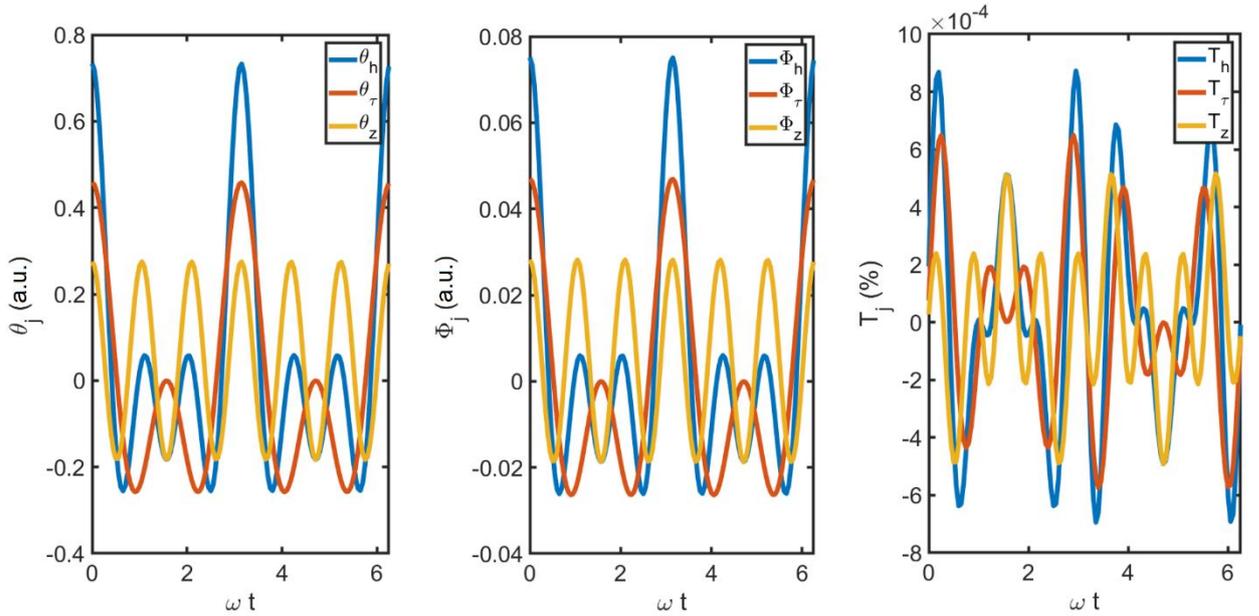

Figure 3. Contributions of the in-plane and out-of-plane **h** components +to the value of (a) $\theta$ angle (b) Faraday rotation (c) transmittance of the structure at $\lambda = 782.5$ nm and normal incidence with magnetization determined directly by Eq. (5).

Symmetry of the structure prohibits [24] other odd magnetization effects at normal incidence. However, even magneto-optical effects can arise resulting in the transmittance modulation that is quadratic on the magnetization components. Generally, at normal incidence of light the transmittance of the magnetic structure can be described as:

$$T = T_0 + \delta_{xx} m_x^2 + \delta_{yy} m_y^2 + \delta_{zz} m_z^2, \tag{11}$$

where the coefficients $\delta_{jj}$ refer to the corresponding even magneto-optical effect produced by the corresponding $m_j = M_j/M_s$ components. Formally, cross-components $\delta_{i \neq j} m_i m_j$ should also be present in Eq. (11), however the numerical calculations show that they have several orders lower values than all the other components in both smooth and perforated films. Eq. (11) can be understood in a sense of the Voigt effect for $m_{x,y}$ components: as the magnetic birefringence arises [25], it results in a small quadratic modification of the Fresnel coefficients leading to the magneto-optical variation of transmittance. The same explanation is valid for the $m_z$ component: as the refractive index of the mode acquires magneto-optical variations, the Fresnel coefficients are modified, too.

The ratio between $\delta_{jj}$ coefficients depend on the magneto-optical properties of the structure. For a smooth iron-garnet film of 30 nm thickness the coefficients are $\delta_{xx} = 2 \cdot 10^{-4}\%$, $\delta_{yy} = 1 \cdot 10^{-4}\%$, $\delta_{zz} = 4 \cdot 10^{-4}\%$. Fig. 2(d) shows that in the considered nanostructure this coefficients are small, too, and this even in magnetization impact could be neglected compared to the Faraday rotation. However, for the nanostructured materials the ratio between these components significantly varies depending on the type of the structure, character of the modes excited in it and so on. For example, in several previous studies [21,26] the plasmonic structure with iron-garnet was designed with $\delta_{yy}$ that prevailed over all other coefficients. Let us analyze the situation where the $\delta_{zz}$ is higher and only the intensity of the transmitted light is measured.

As we are interested in measurement of all three orthogonal components of the weak magnetic field **h**, which are present in $\theta$ temporal dependence as harmonics with different frequencies, let us consider the nanostructure with $\delta_{zz} \gg \delta_{xx}, \delta_{yy}$, which is fulfilled for the considered nanostructure, too. Therefore, only this term in Eq. (11) can be considered further. The temporal dependence of the $m_z$ could be treated as $m_z \approx \theta(t) = \theta_H(t) + \theta_h(t)$, where $\theta_H \gg \theta_h$ are the contributions corresponding to the external magnetic field **H** and the measured **h** in the temporal dependence of $\theta$ (Eq. (5)). Therefore, the transmittance of the structure will experience the following temporal modulation:

$$T = T_0 + \delta_{zz}(\theta_H^2 + 2\theta_H\theta_h), \qquad (12)$$

and according to Eq. (8), the measured magnetic field **h** will be responsible for the following transmittance modulation:

$$T_\tau = h_\tau \frac{K^2}{6A^2H} \delta_{zz} \sin(7\omega t + \varphi_K + \varphi_h) + h_\tau \frac{K^2}{6A^2H} \delta_{zz} \sin(5\omega t - \varphi_K - \varphi_h), \qquad (13)$$

$$T_\perp = h_z \frac{K^2}{6A^2H} \delta_{zz} \sin(9\omega t) - h_z \frac{K^2}{6A^2H} \delta_{zz} \sin(3\omega t), \qquad (14)$$

while the applied external magnetic field **H** will result in modulation at a different frequency

$$T_H = \delta_{zz}(1 - \cos(6\omega t)) \frac{K^2}{18A^2}. \qquad (15)$$

Therefore, Fourier transform of the transmitted through the grating light temporal dependence enables measurement of the in-plane component magnitude and direction, $h_\tau$ and $\varphi_\tau$ at $7\omega, 5\omega$ frequencies, while $3\omega, 9\omega$ components are responsible for the out-of-plane $h_z$ component of the measured magnetic field, correspondingly. Although this method allows one to determine the measured field **h**, its sensitivity is obviously much smaller than the sensitivity of the Faraday method: Eq.c(12) implies that in the considered configuration of the magnetic field the angle associated with the measured magnetic field is multiplied by the $\theta_H$ which is intrinsically small.

Recent studies [23,26] show that the sensitivity of the magneto-optical magnetometers is limited mainly by the optical noise. Analysis using Eq. (9) provides sensitivity for the Faraday effect-based sensor up to 100 pT/Hz$^{1/2}$. Similar calculation using Eqs. (13) and (14) for quadratic magneto-optical intensity effect shows 4 order of magnitude lower value of the sensitivity. However, an advantage of quadratic effect lies in the fact that signal is observed on 7$^{th}$ and 9$^{th}$ harmonics of the magnetic field rotation frequency. It is known that the optical noise quickly decreases at frequencies over the bandwidth of the laser source. For example, an optical noise of a laser with 1 MHz bandwidth more than 4 order of magnitude lower at 10 MHz frequency. It means that application of narrow band laser and increasing of frequency of magnetic field rotation $\omega$ up to some megahertz can provide better sensitivity for quadratic magneto-optical effect with respect to the Faraday effect.

## Summary

A 1D nanograting deposited on an iron-garnet film with strong magnetic anisotropy is promising for the magneto-optical magnetometry. The magnetic properties of such grating enable the magneto-optical measurement of all orthogonal components of the weak magnetic field vector due to the multifold enhancement of the Faraday effect. The 1D grating magneto-optical magnetometer based on measurement of the Faraday rotation was predicted to have 100 pT/Hz$^{1/2}$ sensitivity. It is important that the sensitivity of the magneto-optical signal to all of the spatial components of the measured magnetic field is the same in this scheme.

A second approach, based on the magneto-optical intensity modulation measurement due to the even in magnetization components magneto-optical effects was considered and showed a 4 order of magnitude lower sensitivity.

Excitation of the guided modes allowed us to increase the thin-film Faraday rotation 100 times compared to a smooth film of the same thickness. Although the achieved enhancement is associated with the lower transmittance of the structure, we show that magneto-optical figure of merit and signal-to-noise ratio are 10 times enhanced compared to the uncoated film. Utilization of the thin films instead of the bulk crystals is important for applications, since it allows for significant miniaturization of the device and better spatial resolution.

## Acknowledgements

This work was financially supported by Russian Foundation for Basic Research, RFBR project N 18-29-02120.

## References

1. Kwong, J. S.; Leithauser, B.; Park, J. W.; Yu, C. M. Diagnostic¨ value of magnetocardiography in coronary artery disease and cardiac arrhythmias: a review of clinical data. Int. J. Cardiol. 2013, 167 (5), 1835−1842.

2. Shiono, J.; Horigome, H.; Matsui, A.; Terada, Y.; Miyashita, T.; Tsukada, K. Detection of repolarization abnormalities in patients with cardiomyopathy using current vector mapping technique on magnetocardiogram. International Journal of Cardiovascular Imaging (formerly Cardiac Imaging) 2003, 19 (2), 163−170.

3. Baillet, S. Magnetoencephalography for brain electrophysiology and imaging. Nat. Neurosci. 2017, 20 (3), 327.

4. Espy, M.; Flynn, M.; Gomez, J.; Hanson, C.; Kraus, R.; Magnelind, P.; Peters, M. Ultra-low-field MRI for the detection of liquid explosives. Supercond. Sci. Technol. 2010, 23 (3), No. 034023.

5. Burch, J. L.; Torbert, R. B.; Phan, T. D.; Chen, L. J.; Moore, T. E.; Ergun, R. E.; Wang, S. Electron-scale measurements of magnetic reconnection in space. Science 2016, 352 (6290), aaf2939.

6. Primdahl, F. The fluxgate magnetometer. J. Phys. E: Sci. Instrum. 1979, 12 (4), 241.

7. Vazquez, M.; Knobel, M.; Sánchez, M. L.; Valenzuela, R.; Zhukov, A. P. Giant magnetoimpedance effect in soft magnetic wires for sensor applications. Sens. Actuators, A 1997, 59 (1−3), 20−29.

8. Korepanov, V.; Marusenkov, A. Flux-gate magnetometers design peculiarities. Surveys in geophysics 2012, 33 (5), 1059−1079.

9. Vetoshko, P. M.; Gusev, N. A.; Chepurnova, D. A.; Samoilova, E. V.; Syvorotka, I. I.; Syvorotka, I. M.; Zvezdin, A. K.; Korotaeva, A. A.; Belotelov, V. I. Flux-gate magnetic field sensor based on yttrium iron garnet films for magnetocardiography investigations. Technical Physics Letters 2016, 42, 860–864

10. Goto, T., Dorofeenko, A. V., Merzlikin, A. M., Baryshev, A. V., Vinogradov, A. P., Inoue, M., Lisyansky, AA., Granovsky, A. B. Optical Tamm states in one-dimensional magnetophotonic structures. Physical Review Letters 2008, 101(11), 113902.

11. Kahl, S., Grishin, A. M. Magneto-optical rotation of a one-dimensional all-garnet photonic crystal in transmission and reflection. Physical Review B 2005, 71(20), 205110.

12. Musorin, A. I., Sharipova, M. I., Dolgova, T. V., Inoue, M., Fedyanin, A. A. Ultrafast Faraday rotation of slow light. Physical Review Applied 2016, 6(2), 024012.

13. Borovkova, O. V., Ignatyeva, D. O., Sekatskii, S. K., Karabchevsky, A., Belotelov, V. I. High-Q surface electromagnetic wave resonance excitation in magnetophotonic crystals for supersensitive detection of weak light absorption in the near-infrared. Photonics Research 2020, 8(1), 57-64.

14. Lodewijks, K., Maccaferri, N., Pakizeh, T., Dumas, R. K., Zubritskaya, I., Åkerman, J., Vavassori, P., Dmitriev, A. Magnetoplasmonic design rules for active magneto-optics. Nano Letters 2014, 14(12), 7207-7214.

15. Caballero, B., García-Martín, A., Cuevas, J. C. Hybrid magnetoplasmonic crystals boost the performance of nanohole arrays as plasmonic sensors. ACS Photonics 2016, 3(2), 203-

208.

16. Ignatyeva, D. O., Davies, C. S., Sylgacheva, D. A., Tsukamoto, A., Yoshikawa, H., Kapralov, P. O., Kirilyuk, A., Belotelov, V. I., Kimel, A. V. Plasmonic layer-selective all-optical switching of magnetization with nanometer resolution. Nature Communications 2019, 10(1), 4786.

17. Li, R., Levy, M. Bragg grating magnetic photonic crystal waveguides. Applied Physics Letters 2005, 86(25), 251102.

18. Royer, F., Varghese, B., Gamet, E., Neveu, S., Jourlin, Y., Jamon, D. Enhancement of Both Faraday and Kerr Effects with an All-Dielectric Grating Based on a Magneto-Optical Nanocomposite Material. ACS Omega 2020, 5(6), 2886-2892.

19. Voronov, A. A., Karki, D., Ignatyeva, D. O., Kozhaev, M. A., Levy, M., Belotelov. V. I. Magneto-optics of subwavelength all-dielectric gratings. Optics Express 2020, 28, 17988-17996.

20. Chernov, A. I., Kozhaev, M. A., Ignatyeva, D. O., Beginin, E. N., Sadovnikov, A. V., Voronov, A. A., Karki, D., Levy, M., Belotelov, V. I. All-Dielectric Nanophotonics Enables Tunable Excitation of the Exchange Spin Waves. Nano Lett. 2020, 20(7), 5259–5266.

21. Belotelov, V. I., Kreilkamp, L. E., Akimov, I. A., Kalish, A. N., Bykov, D. A., Kasture, S., Yallapragada, V. J., Achanta, V. G., Grishin, A. M., Khartsev, S. I., Nur-E-Alam, M., Vasiliev, M., Doskolovich, L. L., Yakovlev, D. R., Alameh, K., Zvezdin, A. K., Bayer, M. Plasmon-mediated magneto-optical transparency. Nature Communications 2013, 4, 2128–2128.

22. Syvorotka, I. I., Vetoshko, P. M., Skidanov, V. A., Shavrov, V. G., Syvorotka, I. M., In-Plane Transverse Susceptibility of (111)-Oriented Iron Garnet Films. IEEE Transactions on Magnetics 2015, 51(1), 2000703.

23. Rogachev, A. E., Vetoshko, P. M., Gusev, N. A., Kozhaev, M. A., Prokopov, A. R., Popov, V. V., Dodonov, D. V., Shumilov, A. G., Shaposhnikov, A. N., Berzhansky, V. N., Zvezdin, A. K., Belotelov, V. I. Vector magneto-optical sensor based on transparent magnetic films with cubic crystallographic symmetry. Applied Physics Letters 2016, 109(16), 162403.

24. Borovkova, O. V., Hashim, H., Ignatyeva, D. O., Kozhaev, M. A., Kalish, A. N., Dagesyan, S. A., Shaposhnikov, A. N., Berzhansky, V. N., Achanta, V. G., Panina, L. V., Zvezdin, A. K., Belotelov, V. I. Magnetoplasmonic structures with broken spatial symmetry for light control at normal incidence. Phys. Rev. B 2020, 102, 081405(R).

25. Zvezdin, A. K., Kotov, V. A. Modern magnetooptics and magnetooptical materials. CRC Press, 1997.

26. Knyazev, G. A., Kapralov, P. O., Gusev, N. A., Kalish, A. N., Vetoshko, P. M., Dagesyan, S. A., Shaposhnikov, A. N., Prokopov, A. R., Berzhansky, V. N., Zvezdin, A. K.,

Belotelov, V. I. Magnetoplasmonic crystals for highly sensitive magnetometry. ACS Photonics 2018, 5(12), 4951–4959.